# Opto-Electronic Clock Regeneration - A Tutorial

Palle Jeppesen and Bjarne Tromborg

December 15, 2024

DTU Electro, Technical University of Denmark, Ørsteds Plads, 343, Kgs. Lyngby, 2800, Denmark

**Abstract:** A tutorial on opto-electronic clock regeneration at very high bit rates beyond reach with purely electronic solutions is given. Emphasis is placed on sum frequency generation in a nonlinear $\chi^{(2)}$- material such as $LiNbO_3$. We first provide a basic introduction to CR (clock recovery) and a PLL (phase-locked loop); two examples are considered, an input signal frequency step and a slow input signal frequency. Next we discuss opto-electronic clock recovery based on an OPLL (opto-electronic PLL). The OPLL contains a phase comparator consisting of a planar $LiNbO_3$ waveguide, a lowpass filter, a VCO (voltage controlled oscillator) and a local oscillator laser. The error signal from the comparator determined by the difference in electrical phase between the signal and the VCO controls the VCO. The VCO has two outputs; one that modulates the local oscillator laser and another that triggers a decision circuit that samples the output from the OPLL. The VCO is continuously adjusted by the OPLL so that it will ensure sampling of the signal in the optimal moments.

The theory for sum frequency generation in the $LiNbO_3$ waveguide is treated by introducing a nonlinear optical coefficient in the wave equation for the electrical field at the optical sum frequency generated by the input signal wave and the wave from a local oscillator laser. For monochromatic waves the output intensity is calculated versus the waveguide length with optical phase mismatch as parameter. Ideal phase match, i.e. zero phase mismatch, gives the highest output. If that condition cannot be satisfied, quasi phase matching might be used where a spatially modulated nonlinear optical coefficient versus distance is introduced by periodical poling of the waveguide. This gives an improvement but does not supersede ideal phase match. For time dependent envelopes a kind of transfer function for the envelope of the output field is derived. It relates the Fourier transforms of the envelopes of the signal and clock to the Fourier transform of the output envelope at the sum frequency. The transfer function takes into account the walk-off between the signal and generated wave due to a difference in group velocity; this walk-off effect reduces the output. The transfer function is a sinc-shaped function of the frequency spacing between the optical carrier frequencies of the signal and clock. For two $LiNbO_3$ waveguides with lengths of 30 mm and 60 mm the first zeros occur for 111 GHz and 55 GHz respectively.

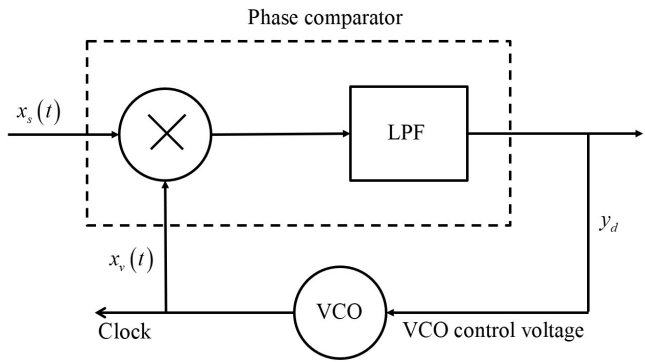

Figure 1: Block diagram of phase-lock loop. VCO: voltage-controlled oscillator. LPF: lowpass filter.

The mathematical foundation for an example of a clock extraction experiment in [5] is given. The input signal had a sinusoidal amplitude modulation at the modulation frequency of 40 GHz. The clock had a fundamental frequency at 10 GHz and a content of higher harmonics. The purpose was to extract a clock at 10 GHz. For experimental validation the 40 GHz signal was observed on an optical sampling scope; the scope was triggered by the recovered clock and for comparison by the frequency synthesizer that was driving the signal source. No noticeable difference between the two signal traces was observed thus demonstrating good quality of the recovered clock. In another experiment in [5] the error signal's sinusoidal dependence of the phase error in frequency lock operation was confirmed; this was done by measuring the power of the error signal versus the phase difference between the signal and clock. Furthermore, clock recovery was demonstrated experimentally at 160 Gbit/s in [5], at 320 Gbit/s in [6] and at 640 Gbit/s in [7, 8].

# 1 Introduction

The purpose of this paper is to present the theory for opto-electronic clock regeneration in very high speed optical communication systems. Digital receivers need CR (clock recovery) which is the process of extracting timing information from a serial data stream to allow the receiving circuit to decode the transmitted symbols. By clock is meant a periodic signal running at a fundamental frequency equal to or very close to the transmitted data rate. For sinusoidal or nearly sinusoidal clocks the clock is generated by a VCO (voltage controlled oscillator) in the receiver and as hinted it is closely aligned to the signal data stream. The clock is used to trigger a sampling circuit. In front of the sampling circuit is placed a differential amplifier in the receiver that switches to high or low output when its input goes above or below a decision threshold; in this way the high or low levels in the bit stream are regenerated but at somewhat random instants. So re-timing is needed and that is provided by a subsequent sampling circuit, that is driven by the clock and which samples the regenerated levels in the optimal moment in the bit interval. The clock helps remove timing jitter in the received signal and from the level regeneration. If a sampled value is above the decision threshold in a decision circuit a 1-bit is decided, if below a 0-bit. In this way CDR (clock and data recovery) is obtained. In systems containing a chain of repeaters each repeater provides 3R-regeneration (re-amplification, re-shaping, re-timing).

In this tutorial, we first provide a basic introduction to CR and PLL (phase-locked loop) that is general in the sense that it covers both electrical and optical communication. To illustrate how the PLL works we consider as examples an input frequency step and a slow input signal variation. Next we treat opto-electronic clock regeneration used in digital optical receivers at such high bit rates that the purely electronic solutions cannot work so opto-electronic solutions are needed. Such solutions involve an OPLL (opto-electronic PLL) which in turn includes a nonlinear $LiNbO_3$ crystal that provides sum frequency generation between an input optical signal and a laser. A basic description of the sum frequency generation is given. A general nonlinear wave equation is derived where the detailed steps are given in an appendix. The wave equation is used to derive the output intensity. The first case we consider is two monochromatic input waves and we derive the output intensity for ideal phase match which gives the highest output. If that condition is not possible to obtain quasi phase matching can be used where spatial modulation of the nonlinear coefficient is introduced. The other case we consider is two input waves with time dependent envelopes; for that case the theory is verified experimentally for a 40 GHz signal and a 10 GHz clock in [5].

## 2 Clock recovery and PLL basics

The clock is controlled in frequency and phase by a PLL and the goal is to align the clock to the signal data rate. A block diagram of a PLL is shown in Figure (1) [1, 2, 3, 4]. It consists of a phase comparator and a VCO. The phase comparator consists of a multiplier followed by a LPF (lowpass filter). The phase comparator produces a control voltage $y_d(t)$ that is related to the frequency and phase difference between the signal and clock. The VCO output signal $x_v(t)$ is the clock. The PLL is said to achieve lock when the transmitted data rate and clock frequency become equal and the phase difference constant. Before lock the control voltage changes the VCO in frequency and phase so that the PLL achieves lock. The function of the low pass filter is to ensure there are no disturbing high frequency components in $y_d(t)$.

Let us now discuss a simple mathematical model of the clock recovery process. The input signal $x_s(t)$ to the PLL is given by

$$x_s(t) = A_s \cos[\theta_s(t)] \tag{1}$$

where $A_s$ is the amplitude, $\theta_s(t) = \omega_s t + \phi_s(t)$, $\omega_s$ the angular frequency and $\phi_s(t)$ the phase; $\phi_s(t)$ may have small variations. The VCO is noise and jitter free and its output is given by

$$x_v(t) = A_v \cos[\theta_v(t)] \tag{2}$$

where $A_v$ is the amplitude, $\theta_v(t) = \omega_v t + \phi_v(t) + \pi/2$, $\omega_v$ is the VCO's free running angular frequency and $\phi_v(t)$ is the phase given by

$$\phi_v(t) = \phi_v(0) + K_v \int_0^t y_d(t')\, dt'. \tag{3}$$

Here $K_v$ is the sensitivity of the VCO, $y_d(t)$ is the control voltage obtained by lowpass filtering of the product of $x_s(t)$ and $x_v(t)$, i.e.

$$y_d(t) = \langle x_s(t) \cdot x_v(t) \rangle_{LP} = K_a \sin[\Delta\omega t + \phi_s(t) - \phi_v(t)] = K_a \sin[\epsilon(t)] \tag{4}$$

where $K_a = \frac{1}{2} A_s A_v$, $\Delta\omega = \omega_s - \omega_v$ and the angular error $\epsilon(t)$ is

$$\epsilon(t) = \theta_s - \theta_v + \frac{\pi}{2} = \Delta\omega t + \phi_s(t) - \phi_v(t). \tag{5}$$

Note that the control voltage is called the error signal in Section 5. The subscript $s$ refers to signal, $v$ to VCO and $d$ to difference in phase. The $\pi/2$ term in $\theta_v(t)$ serves the purpose of obtaining the sinusoidal dependence in (4) so that zero control voltage corresponds to zero angular error.

As already mentioned the phase $\phi_s$ may have small variations and it is the purpose of the PLL to extract the angular signal frequency $\omega_s$. Note that (3) reflects the key feature of a VCO, namely that its frequency and phase are adjusted by a control voltage.

Now from (3) and (4) we obtain

$$\dot{\phi}_v(t) = K_v y_d(t) = K \sin[\epsilon(t)] \tag{6}$$

where the loop gain is $K = K_v K_a$ and where $\tau = 1/K$ is a time constant to be used later. Then from (5) and (6) we obtain the fundamental equation

$$\dot{\epsilon}(t) + K \sin[\epsilon(t)] = \Delta\omega + \dot{\phi}_s(t) \tag{7}$$

where $\Delta\omega$ is assumed constant.

The relation between $\dot{\epsilon}(t)$ and $\epsilon(t)$ is shown in Figure 2 for an example where $K > \Delta\omega + \dot{\phi}_s(t) \geq 0$. Here $\dot{\epsilon}(t)$ is zero at the two points $\epsilon = \epsilon_e$ and $\epsilon = \pi - \epsilon_e$ given by

$$K \sin[\epsilon_e(t)] = \Delta\omega + \dot{\phi}_s(t). \tag{8}$$

In the $\epsilon$-intervals where $\dot{\epsilon}(t)$ is positive the value of $\epsilon(t)$ will increase with time and correspondingly, the value of $\epsilon(t)$ decreases with time where $\dot{\epsilon}(t)$ is negative.

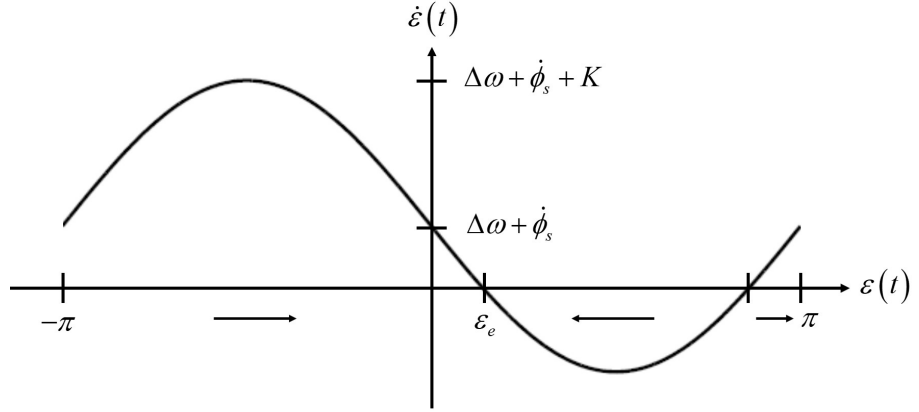

Figure 2: The curve shows $\dot{\epsilon}(t) = -K\sin[\epsilon(t)] + \Delta\omega + \dot{\phi}_s(t)$. The arrows show the direction of the drift of $\epsilon(t)$ in the intervals $[-\pi, \epsilon_e]$, $[\epsilon_e, \pi - \epsilon_e]$ and $[\pi - \epsilon_e, \pi]$. [3, 4]

The direction of the drift of $\epsilon(t)$ is indicated by arrows in the intervals. For constant $\Delta\omega + \dot{\phi}_s$ it shows that the point $\epsilon = \epsilon_e = \arcsin\left[\left(\Delta\omega + \dot{\phi}_s\right)\tau\right]$ is a stable stationary solution and the point $\epsilon = \pi - \epsilon_e$ is an unstable stationary solution. When $\epsilon(t)$ has drifted to the stable point $\epsilon_e$, the PLL is described as being phase-locked and $\epsilon(t)$ remains fixed at the static error.

In the linear regime where $\sin[\epsilon(t)] \simeq \epsilon(t)$ we can approximate (7) by

$$\dot{\epsilon}(t) + \frac{1}{\tau}\epsilon(t) = \Delta\omega + \dot{\phi}_s(t) \ . \tag{9}$$

Eq. (9) has the solution

$$\epsilon(t) = \epsilon(0)e^{-t/\tau} + \int_0^t e^{-(t-t')/\tau}\left(\Delta\omega + \dot{\phi}_s(t')\right)dt' \ . \tag{10}$$

However, it is only an approximate solution to (7) if $\epsilon(t)$ is all the time in the linear regime, i.e. $|\epsilon(t)| < \pi/4$. For constant $\dot{\phi}_s(t) = \dot{\phi}_s$ the result $\epsilon(t) \simeq \epsilon(0)e^{-t/\tau} + \tau(\Delta\omega + \dot{\phi}_s)(1 - e^{-t/\tau})$ confirms that $\epsilon(t)$ approaches $\epsilon_e = \left(\Delta\omega + \dot{\phi}_s\right)\tau$ given by (8) provided $\epsilon_e$ is in the linear regime.

For $t \gg \tau$ we see that

$$\begin{aligned}\epsilon(t) &\simeq \tau\Delta\omega + \int_0^t e^{-(t-t')/\tau}\dot{\phi}_s(t')\,dt' \\ &\simeq \tau\Delta\omega + \phi_s(t) - \langle\phi_s(t)\rangle\end{aligned} \tag{11}$$

where
$$\langle \phi_s(t) \rangle = \frac{1}{\tau} \int_0^t \phi_s(t') e^{-(t-t')/\tau} dt' \qquad (12)$$
is a sort of average or lowpass filtered version of the input phase $\phi_s(t)$. Hence from (5) and recalling $\theta_s(t) = \omega_s t + \phi_s(t)$

$$\begin{aligned}\theta_v(t) &= \theta_s(t) + \frac{\pi}{2} - \epsilon(t) \\ &\simeq \omega_s t + \langle \phi_s(t) \rangle - \Delta\omega\tau + \frac{\pi}{2}.\end{aligned} \qquad (13)$$

This means the VCO oscillates at the same angular frequency $\omega_s$ as the signal. In other words the PLL is in lock-in operation and has recovered the signal frequency. The phase of the VCO is $\langle \phi_s(t) \rangle - \Delta\omega\tau + \pi/2$. For slow phase fluctuations compared to $\tau$ we have $\langle \phi_s(t) \rangle \simeq \phi_s(t)$ so the VCO will get the phase $\phi_v(t) \simeq \phi_s(t) - \Delta\omega\tau$, i.e. the VCO will track the signal also in phase except for the phase delay $\Delta\omega\tau$.

For $\tau \left| \Delta\omega + \dot\phi_s(t) \right| > 1$ the sinusoidal curve in Figure 2 will not intersect the $\epsilon$-axis and $\epsilon(t)$ will keep drifting to the right or left. In this case phase-locking is therefore not possible.

In order to further illustrate some of the implications of (10) let us consider two thought experiments.

## 3   Input frequency step

For $t \leq 0$ assume the PLL is in lock-in operation, $\omega_s$ and $\omega_v$ are constants and equal and therefore $\Delta\omega = 0$; also $\phi_s(t)$ and $\phi_v(t)$ are constants but not necessarily equal. Suppose that at $t = 0^+$ the input signal changes in frequency by the step $\delta\omega$, i.e. $\dot\phi_s(t) = \delta\omega$ and hence $\phi_s(t) = \delta\omega \cdot t + \phi_s(0)$. We assume $\delta\omega$ is positive and constant. For $t > 0$ we get from (12)

$$\begin{aligned}\langle \phi_s(t) \rangle &= \frac{1}{\tau} \int_0^t (\phi_s(0) + \delta\omega \cdot t') e^{-(t-t')/\tau} dt' \\ &= \phi_s(0)\left(1 - e^{-t/\tau}\right) + \delta\omega\left(t - \tau + \tau e^{-t/\tau}\right).\end{aligned} \qquad (14)$$

Hence for $t \gg \tau$
$$\langle \phi_s(t) \rangle \simeq \phi_s(0) + \delta\omega(t - \tau) \qquad (15)$$
and then from (13)

$$\begin{aligned}\theta_v(t) &\simeq \omega_s t + \phi_s(0) + \delta\omega(t - \tau) + \frac{\pi}{2} \\ &= (\omega_s + \delta\omega)t + \phi_s(0) - \delta\omega \cdot \tau + \frac{\pi}{2}.\end{aligned} \qquad (16)$$

The equation shows that for $t \gg \tau$ the PLL automatically reaches a new equilibrium, i.e. a new lock-in operation, where the VCO has the same frequency as the signal, while its phase has changed from the initial value $\phi_v(0) + \pi/2$ to $\phi_s(0) - \delta\omega \cdot \tau + \pi/2$.

# 4 Slow signal variations and PLL bandwidth

For $t \leq 0$ we assume the PLL is in the same condition as in the previous example. For $t > 0$, $\phi_s(t)$ is changed to $\phi_s(t) = \phi_s(0) + \delta\phi \sin(\omega_m t)$ where $\delta\phi$ is constant and $\omega_m$ is an angular modulation frequency.

For $t > 0$ we find by (12)

$$\langle \phi_s(t) \rangle = \frac{1}{\tau} \int_0^t \{\phi_s(0) + \delta\phi \sin(\omega_m t')\} e^{-(t-t')/\tau} dt'$$
$$= \phi_s(0)\left(1 - e^{-t/\tau}\right) + \frac{\delta\phi}{1 + (\omega_m \tau)^2}\left[\omega_m \tau e^{-t/\tau} + \sin(\omega_m t) - \omega_m \tau \cos(\omega_m t)\right] \quad (17)$$

so for $t \gg \tau$

$$\langle \phi_s(t) \rangle \simeq \phi_s(0) + \frac{\delta\phi}{1 + (\omega_m \tau)^2}\left[\sin(\omega_m t) - \omega_m \tau \cos(\omega_m t)\right]$$
$$= \phi_s(0) + \frac{\delta\phi}{\sqrt{1 + (\omega_m \tau)^2}} \sin(\omega_m t - \psi) \quad (18)$$

where $\tan(\psi) = \omega_m \tau$. Inserted in (13) this gives

$$\theta_v(t) \simeq \omega_s t + \phi_s(0) + \frac{\delta\phi}{\sqrt{1 + (\omega_m \tau)^2}} \sin(\omega_m t - \psi) + \frac{\pi}{2}. \quad (19)$$

For $\omega_m \ll 1/\tau$, (19) simplifies to $\theta_v(t) \simeq \omega_s t + \phi_s(0) + \delta\phi \sin(\omega_m t - \psi) + \pi/2$. This result shows the VCO has the same frequency as the signal and hence provides clock recovery. There is a small phase delay $\psi$ in the phase modulation, but that has no consequence for the clock recovery. The modulation frequency can be claimed to represent any slow frequency variation; hence the PLL provides clock recovery for random slow phase variations. The amplitude of the phase modulation in (19) has decreased with a factor of $\sqrt{2}$ at $\omega_m = 1/\tau$ which then can be considered the angular bandwidth. The PLL has performed low pass filtering of the phase modulation.

For $\omega_m \gg 1/\tau$ we see $\theta_v(t) \simeq \omega_s t + \phi_s(0) + \pi/2$ which means the clock has the same frequency as the signal and the high frequency phase modulation has been eliminated. Clock recovery has been obtained. In general high frequency phase modulation above the loop bandwidth is suppressed or eliminated.

# 5 Introduction to optical clock regeneration and optical phase-locked loop

In this section we will discuss optical clock regeneration used in digital optical receivers at such high bit rates that the purely electronic solutions cannot work so opto-electronic solutions are needed. For example, such a technique was demonstrated in 640 Gbit/s receivers in [5, 7, 8]. The principle is shown in Figure 3.

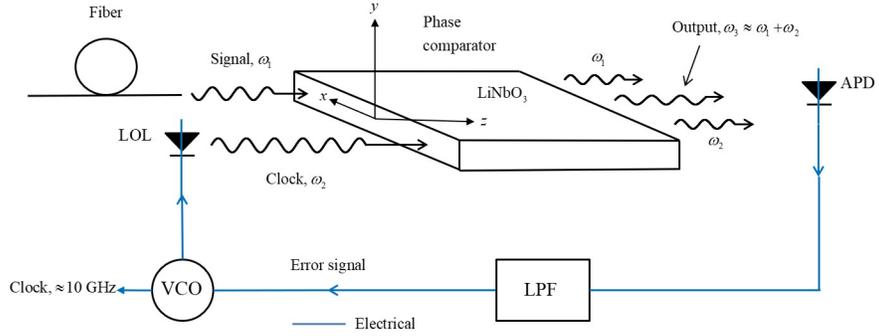

Figure 3: Principle for sum frequency generation, phase comparison and clock recovery. VCO: voltage controlled oscillator. LOL: local oscillator laser. APD: avalanche photodiode. LPF: lowpass filter.

The clock is controlled by an OPLL. In the first experiment [5] we will refer to later the fundamental frequency of the clock is 10 GHz and the input is a sinusoidal signal amplitude modulated at 40 GHz; thereafter we will briefly refer to another experiment [7] that shows that the method described can also work in a 640 Gbit/s OTDM (optical time division multiplex) system.

With reference to Figure 3 the received signal from a fiber and the beam from a LOL (local oscillator laser) are both injected into a phase comparator consisting of a planar waveguide made of nonlinear $\chi^{(2)}$−material such as LiNbO$_3$ [9]. An output beam at the sum frequency is detected in a Si APD (avalanche photodiode) which is blind to irrelevant other outputs at higher optical frequencies. The APD provides inherent lowpass filtering because of its small electrical bandwidth. The photocurrent from the APD is further lowpassed filtered in a subsequent LPF (lowpass filter) and as explained in Section 2 the output is a slow error signal given by the frequency and phase difference between the signal and the clock. The error signal controls a VCO. The VCO has two outputs; one modulates the LOL and the other triggers a decision circuit (not shown) that samples the received signal. The VCO is continuously adjusted by the OPLL so that it will ensure sampling of the signal in the optimal moments.

We want to derive the error signal. But first we give a basic introduction to sum frequency generation by means of two wave mixing in a nonlinear $\chi^{(2)}$−material, and in order to do that we derive a nonlinear equation that is the basis for the further derivations.

# 6 Wave equation for the electric field in the LiNbO$_3$ crystal

Consider the real physical electric field in vectorial form $\boldsymbol{E}$ that propagates in a linear, isotropic, homogeneous, nonconductive, nonmagnetic medium without charges. The electric field is a function of $x, y, z$ and $t$ but for simplicity we suppress this notation. With reference to [10] it can be shown based on Maxwell's equations and constitutive equations that the field propagates according to the vectorial wave equation

$$\nabla^2 \boldsymbol{E} - \mu_0 \varepsilon_0 \left[ \frac{\partial^2 \boldsymbol{E}}{\partial t^2} + \chi(t) \otimes \frac{\partial^2 \boldsymbol{E}}{\partial t^2} \right] = 0. \qquad (20)$$

Here $\varepsilon_0 = 8.854 \cdot 10^{-12}$ F/m is the absolute permittivity of vacuum, $\mu_0 = 4\pi \cdot 10^{-7}$ H/m the absolute permeability of vacuum, $\nabla = \hat{\boldsymbol{x}} \frac{\partial}{\partial x} + \hat{\boldsymbol{y}} \frac{\partial}{\partial y} + \hat{\boldsymbol{z}} \frac{\partial}{\partial x}$ is the nabla operator and $\hat{\boldsymbol{x}}$, $\hat{\boldsymbol{y}}$ and $\hat{\boldsymbol{z}}$ the unit vectors along the three coordinate axes. The symbol $\otimes$ means convolution, i.e. $[a \otimes b](t) = \int_{-\infty}^{\infty} a(t-t') b(t') dt'$ for functions $a(t)$ and $b(t)$. The function $\chi(t)$ is the electric susceptibility response related to the polarization $\boldsymbol{P} = \varepsilon_0 \chi(t) \otimes \boldsymbol{E}$. In the convolution term in (20) the expression $\frac{\partial^2 \boldsymbol{E}}{\partial t^2}$ plays the role of input and $\chi(t)$ the impulse respnse in analogy with an electronic two-port. When the response is infinitely fast, i.e. $\chi(t) = \tilde{\chi}_0 \delta(t)$ where $\tilde{\chi}_0$ is constant, the convolution term simplifies to $\tilde{\chi}_0 \cdot \frac{\partial^2 \boldsymbol{E}}{\partial t^2}$.

# 7 Sum frequency generation in a nonlinear material

For a nonlinear material the nonlinearity is taken into account by introducing a nonlinear susceptibility (see below). Because of the nonlinear characteristics the two input waves (signal and clock) generate new waves including one at the sum frequency.

We will study the sum frequency generation. First step, however, is to derive the nonlinear equation that governs the sum frequency generation. To begin with we work with the real physical fields as opposed to complex fields or envelopes or Fourier transforms. In order to focus on the nonlinear process we will ignore waveguiding effects and work with plane waves although the experiments in [5] were based on planar or waveguide devices. So we consider an infinitely wide plane wave that propagates in a LiNbO$_3$ crystal in the $z$-direction with an electric field that only has an $x$-component with no $x$- or $y$-dependence; in that case $\boldsymbol{E} = \hat{\boldsymbol{x}} E_x(z,t)$ and $\nabla \cdot \boldsymbol{E} = 0$. In the following we suppress the $x$−index and the $z$- and time-dependence. In the previous Section 6 we stated the vectorial wave equation (20) for a linear, charge free and nonconducting material; we now write this equation in scalar form

$$\frac{\partial^2 E}{\partial z^2} - \mu_0 \varepsilon_0 \left[ \frac{\partial^2 E}{\partial t^2} + \chi(t) \otimes \frac{\partial^2 E}{\partial t^2} \right] = 0 \qquad (21)$$

or for $P = \varepsilon_0 \chi \otimes E$

$$\frac{\partial^2 E}{\partial z^2} - \mu_0 \varepsilon_0 \frac{\partial^2 E}{\partial t^2} = \mu_0 \frac{\partial^2 P}{\partial t^2}. \qquad (22)$$

However, for the nonlinear material LiNbO$_3$ we use

$$P = P_L + P_{NL} \qquad (23)$$

where the linear part is
$$P_L = \varepsilon_0 \chi(t) \otimes E \qquad (24)$$
with the linear susceptibility $\chi$, and a nonlinear part is
$$P_{NL} = \varepsilon_0 \chi^{(2)} E^2 = 2dE^2 \qquad (25)$$
where $\chi^{(2)}$ is the 2nd order nonlinear susceptibility and $d = \frac{1}{2}\varepsilon_0\chi^{(2)}$ is the nonlinear optical coefficient [9]. We assume $\chi^{(2)}$ and hence $d$ are non-dispersive meaning they are constant in the frequency domain and give instantaneous response in the time domain; they are also real-valued. From (22), (23), (24) and (25) we can now derive
$$\left[\frac{\partial^2}{\partial z^2} - \mu_0\varepsilon_0\frac{\partial^2}{\partial t^2} - \mu_0\varepsilon_0\chi(t)\otimes\frac{\partial^2}{\partial t^2}\right]E = \mu_0\frac{\partial^2 P_{NL}}{\partial t^2}. \qquad (26)$$
Note that to derive (21) $\nabla(\nabla \cdot \boldsymbol{E}) = 0$ was used. This condition is satisfied for the plane wave we are considering. More generally, it is pointed out in [11] that the contribution from the term $\nabla(\nabla \cdot \boldsymbol{E})$ is usually small in nonlinear optics in cases of interest. Now, we will use (26) for the case where wave 1 (signal) and wave 2 (clock) generate a new wave 3 at the sum frequency to be used as the error signal in the LiNbO$_3$ crystal; therefore for the total field we set $E = E_1 + E_2 + E_3$ and for the nonlinear polarization we get

$$P_{NL} = \varepsilon_0\chi^{(2)}E^2 = 2dE^2 = 2d(E_1 + E_2 + E_3)^2$$
$$= 2d\left(E_1^2 + E_2^2 + E_3^2 + 2E_1E_2 + 2E_1E_3 + 2E_2E_3\right). \qquad (27)$$

Eq. (26) can then be written

$$\left[\frac{\partial^2}{\partial z^2} - \mu_0\varepsilon_0\frac{\partial^2}{\partial t^2} - \mu_0\varepsilon_0\chi(t)\otimes\frac{\partial^2}{\partial t^2}\right](E_1 + E_2 + E_3)$$
$$= \mu_0 2d\frac{\partial^2}{\partial t^2}\left(E_1^2 + E_2^2 + E_3^2 + 2E_1E_2 + 2E_1E_3 + 2E_2E_3\right). \qquad (28)$$

This is the wave equation we shall use in the following where we focus on the term oscillating at the angular sum frequency $\omega_3 = \omega_1 + \omega_2$.

## 8 Sum frequency generation based on monochromatic waves

In order to start with a simple case we assume two input waves 1 and 2 that are monochromatic and generate a new wave 3 which is also monochromatic. Furthermore, we assume waves 1 and 2 are strong and only loose little power when generating wave 3 and therefore propagate with constant amplitudes. We write the electric fields in the form

$$E_k(z,t) = \frac{1}{2}\left[A_k(z)e^{j(\omega_k t - \beta_{0k}z)} + c.c.\right] \tag{29}$$

for $k = 1, 2, 3$. Here $A_1(z) = A_1$ and $A_2(z) = A_2$ are constant envelopes. The angular frequencies are called $\omega_k$. The propagation constants are given by $\beta_{0k} = \beta(\omega_k)$ where

$$\beta^2(\omega) = \frac{\omega^2}{c^2}\left(1 + \tilde{\chi}(\omega)\right). \tag{30}$$

$c = 1/\sqrt{\mu_0 \varepsilon_0}$ is the velocity of light in vacuum, and $\tilde{\chi}(\omega)$ is the Fourier transform

$$\tilde{\chi}(\omega) = \int_{-\infty}^{\infty} \chi(t)e^{-j\omega t}dt. \tag{31}$$

The refractive index $n_k$ at angular frequency $\omega_k$ is $n_k = \sqrt{1 + \tilde{\chi}(\omega_k)}$ so $\beta_{0k} = \omega_k n_k/c$. We ignore loss by absorption and assume $\beta_{0k}$ is real. Furthermore, $A_3(z)$ is assumed to be slowly varying compared to $e^{-j\beta_{03}z}$.

In (28) the term $2E_1 E_2$ is given by

$$2E_1 E_2 = \frac{1}{2}\left\{A_1 A_2 e^{j[(\omega_1+\omega_2)t - (\beta_{01}+\beta_{02})z]} + A_1 A_2^* e^{j[(\omega_1-\omega_2)t - (\beta_{01}-\beta_{02})z]} + c.c.\right\}. \tag{32}$$

Here c.c. means the complex conjugate of the previous two terms in the curled bracket; the term $2E_1 E_2$ oscillates at $\omega_1 + \omega_2$ and $|\omega_1 - \omega_2|$. Similarly $2E_1 E_3$ oscillates at $\omega_1 + \omega_3$ and $|\omega_1 - \omega_3|$, and $2E_2 E_3$ at $\omega_2 + \omega_3$ and $|\omega_2 - \omega_3|$. Furthermore, $E_1^2$, $E_2^2$ and $E_3^2$ contain dc terms and terms that oscillate at $2\omega_1$, $2\omega_2$ and $2\omega_3$ respectively. The only term that can drive a new wave synchronously at the angular sum frequency $\omega_3 = \omega_1 + \omega_2$ is $2E_1 E_2$. With reference to (28) we therefore focus on the equation

$$\left[\frac{\partial^2}{\partial z^2} - \mu_0 \varepsilon_0 \frac{\partial^2}{\partial t^2} - \mu_0 \varepsilon_0 \chi(t) \otimes \frac{\partial^2}{\partial t^2}\right]\frac{1}{2}\left[A_3(z)e^{j(\omega_3 t - \beta_{03}z)} + c.c.\right]$$
$$= \mu_0 d \frac{\partial^2}{\partial t^2}\left[A_1 A_2 e^{j[(\omega_1+\omega_2)t - (\beta_{01}+\beta_{02})z]} + c.c.\right]. \tag{33}$$

Using the slowly varying envelope approximation $\frac{d^2 A_3(z)}{dz^2} \simeq 0$ and $\beta_{03} = \omega_3 \sqrt{1 + \tilde{\chi}(\omega_3)}/c$ we find

$$\left[\frac{\partial^2}{\partial z^2} - \mu_0 \varepsilon_0 \frac{\partial^2}{\partial t^2} - \mu_0 \varepsilon_0 \chi(t) \otimes \frac{\partial^2}{\partial t^2}\right]\frac{1}{2}A_3(z)e^{j(\omega_3 t - \beta_{03}z)} \simeq -j\beta_{03}\frac{dA_3(z)}{dz}e^{j(\omega_3 t - \beta_{03}z)}. \tag{34}$$

Inserting this result in (33) gives

$$j\beta_{03}\frac{dA_3(z)}{dz}e^{j(\omega_3 t - \beta_{03}z)} + c.c. \simeq \mu_0 d(\omega_1 + \omega_2)^2\left[A_1 A_2 e^{j[(\omega_1+\omega_2)t - (\beta_{01}+\beta_{02})z]} + c.c.\right] \tag{35}$$

which for $\omega_1 + \omega_2 = \omega_3$ and wave impedance $\eta_3 = \sqrt{\frac{\mu_0}{\varepsilon_0}}/n_3$ leads to

$$\frac{dA_3(z)}{dz} = -j\omega_3 \eta_3 d A_1 A_2 e^{j(\beta_{03}-\beta_{01}-\beta_{02})z}. \tag{36}$$

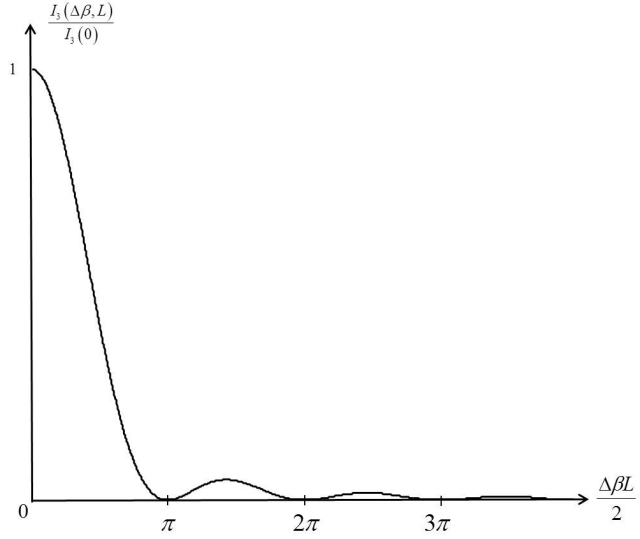

Figure 4: Normalized intensity of wave 3 at the sum angular frequency $\omega_3$ versus the positive half product of phase mismatch and length of crystal.

We now define the phase mismatch $\Delta\beta = \beta_{03} - \beta_{01} - \beta_{02}$ and assume the boundary condition $A_3(0)=0$. For the length $L$ of the LiNbO$_3$ crystal we find the output envelope

$$A_3(L) = -j\omega_3\eta_3 dA_1 A_2 \int_0^L e^{j\Delta\beta z}dz = -j\omega_3\eta_3 dA_1 A_2 L e^{j\frac{\Delta\beta L}{2}} \operatorname{sinc}\left(\frac{\Delta\beta L}{2\pi}\right). \quad (37)$$

In the following we take the wave impedances into account when determining the intensities $I_1$, $I_2$ and $I_3$ of the three waves. We find $I_1 = \frac{1}{2\eta_1}|A_1|^2$, $I_2 = \frac{1}{2\eta_2}|A_2|^2$ and

$$I_3(\Delta\beta, L) = \frac{1}{2\eta_3}|A_3(L)|^2 = 2\eta_1\eta_2\eta_3 I_1 I_2 (\omega_3 dL)^2 \operatorname{sinc}^2\left(\frac{\Delta\beta L}{2\pi}\right). \quad (38)$$

In Figure 4 is shown the normalized intensity $I_3(\Delta\beta, L)/I_3(0, L) = \operatorname{sinc}^2\left(\frac{\Delta\beta L}{2\pi}\right)$ versus $\frac{\Delta\beta L}{2}$; we see a $\operatorname{sinc}^2(x)$ type dependence.

# 9 Ideal phase match

The intensity of wave 3 is maximum when the condition for ideal phase match

$$\Delta\beta = \beta_{03} - \beta_{01} - \beta_{02} = 0 \quad (39)$$

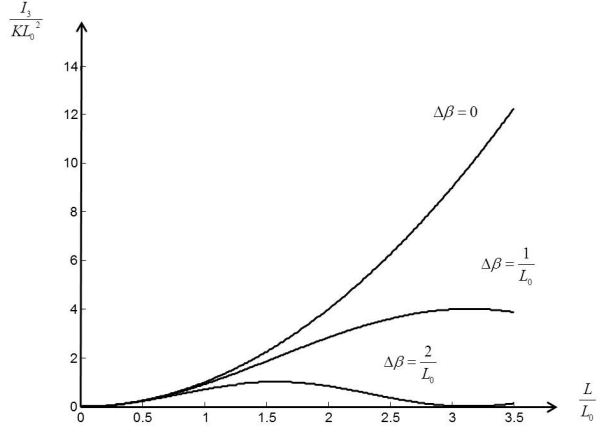

Figure 5: Normalized intensity of wave 3 at sum frequency $\omega_3$ versus normalized length of crystal and with $\Delta\beta$ as parameter.

is satisfied. The first zero occurs for $\frac{\Delta\beta L}{2} = \pi$ which indirectly defines the coherence length $L_c = 2\pi/\Delta\beta$. Defining $K = 2\eta_1\eta_2\eta_3 \left(\omega_3 d\right)^2 I_1 I_2$ we can write

$$I_3\left(\Delta\beta, L\right) = KL^2 \left(\frac{\sin\frac{\Delta\beta L}{2}}{\frac{\Delta\beta L}{2}}\right)^2. \tag{40}$$

If we introduce an arbitrary normalization length $L_0$ in this formula we can for the normalized intensity obtain

$$\frac{I_3\left(\Delta\beta, L\right)}{KL_0^2} = \begin{cases} \left(\frac{L}{L_0}\right)^2 & \text{for } \Delta\beta = 0 \\ 4\sin^2\left(\frac{L}{2L_0}\right) & \text{for } \Delta\beta = \frac{1}{L_0} \\ \sin^2\left(\frac{L}{L_0}\right) & \text{for } \Delta\beta = \frac{2}{L_0} \end{cases} \tag{41}$$

which is plotted in Figure 5 versus $L/L_0$ with $\Delta\beta$ as parameter. Large phase mismatch is seen to cause small output at the sum frequency.

## 10 Quasi phase matching

Because $\beta_{0k} = \omega_k n_k / c$ the ideal phase match condition (39) can also be written $\omega_3 n_3 = \omega_1 n_1 + \omega_2 n_2$. In cases where this condition cannot be satisfied simultaneously with the condition $\omega_3 = \omega_1 + \omega_2$ because the refractive indices vary with frequency quasi phase matching (QPM) can be introduced. For example in periodically poled $LiNbO_3$ it is possible to apply a periodic modulation versus distance to the nonlinear optical coefficient so $d\left(z\right)$ can be expressed by [5, 12]

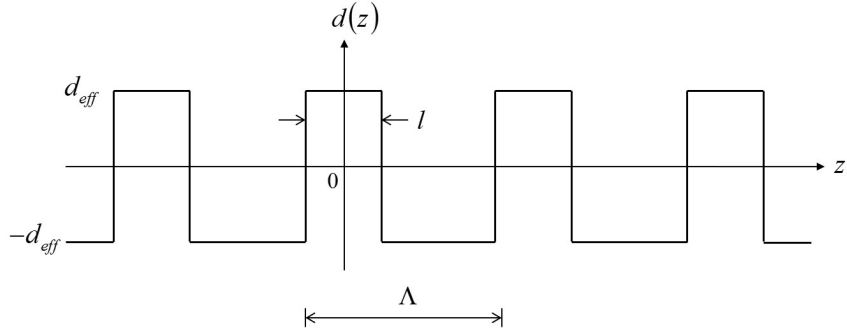

Figure 6: Square wave modulated nonlinear optical coefficient.

$$d(z) = \sum_{m=-\infty}^{\infty} d_m e^{jm 2\pi \frac{z}{\Lambda}} . \qquad (42)$$

Here $\Lambda$ is the period of the spatial modulation, $d_m$ is the Fourier coefficient given by $d_m = \frac{1}{\Lambda} \int_0^{\Lambda} d(z) e^{-jm \frac{2\pi}{\Lambda} z} dz$ and $d_{-m} = d_m^*$ since $d(z)$ is real.

With the modulated coefficient $d(z)$ (36) becomes

$$\frac{dA_3(z)}{dz} = -j\omega_3 \eta_3 A_1 A_2 \sum_{m=-\infty}^{\infty} d_m e^{j\left(m\frac{2\pi}{\Lambda} + \Delta\beta\right)z} . \qquad (43)$$

As an example, consider a grating with "square wave modulation" and where the sign of the optical nonlinear coefficient has been periodically inverted as shown in Figure 6 [5, 12, 13]. The period is $\Lambda$, the width of the region with $d(z) = d_{eff}$ is $l$ and the width with $d(z) = -d_{eff}$ is $\Lambda - l$. The curve is symmetrical around $z = 0$. The Fourier coefficient $d_m$ is given by $d_m = \frac{1}{\Lambda} \int_{-\frac{l}{2}}^{\Lambda - \frac{l}{2}} d(z) e^{-j2\pi m \frac{z}{\Lambda}} dz = \frac{2}{\pi m} d_{eff} \sin\left(\pi m \frac{l}{\Lambda}\right)$ for $m \neq 0$ and $d_0 = d_{eff}(2l - \Lambda)/\Lambda$. For the special case where $l = \frac{\Lambda}{2}$, we find $d_0 = 0$, $d_m = \frac{2}{\pi m} d_{eff} \sin(m\pi/2)$ and

$$\frac{dA_3(z)}{dz} = K_b \sum_{m=1}^{\infty} \frac{1}{m} \sin(\frac{m\pi}{2}) \left(e^{jm\frac{2\pi}{\Lambda}z} + e^{-jm\frac{2\pi}{\Lambda}z}\right) e^{j\Delta\beta z} \qquad (44)$$

where we have defined $K_b = -j\omega_3 \eta_3 A_1 A_2 d_{eff} \frac{2}{\pi}$. For 1st order quasi-phase matching with $\Delta\beta = -\frac{2\pi}{\Lambda}$ the solution to (44) is

$$A_3(L) = K_b L \sum_{m=1}^{\infty} \frac{1}{m} \sin\left(\frac{m\pi}{2}\right) \left[e^{jm\frac{\pi}{\Lambda}L} \operatorname{sinc}\frac{(m-1)L}{\Lambda} + e^{-jm\frac{\pi}{\Lambda}L} \operatorname{sinc}\frac{(m+1)L}{\Lambda}\right] e^{-j\frac{\pi L}{\Lambda}} \qquad (45)$$

assuming $A_3(0) = 0$. The contribution from $m = 1$ is $K_b L \left[1 + e^{-j2\pi \frac{L}{\Lambda}} \operatorname{sinc}\left(\frac{2L}{\Lambda}\right)\right]$ where the second term $\operatorname{sinc}\left(\frac{2L}{\Lambda}\right)$ averages out for $L \gg \Lambda$. Similarly, the sinc functions average out in the summation in (45) for $m > 1$ and $L >> \Lambda$, so $A_3(L) \simeq K_b L$ for large $L$. The corresponding intensity is

$$I_3(L) = \frac{1}{2\eta_3} \left| A_3(L)^2 \right| \simeq \frac{1}{2\eta_3} \left( |K_b| L \right)^2. \tag{46}$$

For ideal phase matching with $\Delta\beta = 0$ the solution to (44) becomes

$$A_3(L) = K_b \frac{\Lambda}{\pi} \sum_{m=1}^{\infty} \frac{1}{m^2} \sin\left(\frac{m\pi}{2}\right) \sin\left(\frac{2\pi L}{\Lambda}\right). \tag{47}$$

The first term gives $A_3(L) = K_b \frac{\Lambda}{\pi} \sin\left(\frac{2\pi L}{\Lambda}\right)$ with intensity

$$I_3(L) = \frac{1}{2\eta_3} \left| A_3(L)^2 \right| = \frac{1}{2\eta_3} \left( |K_b| \frac{\Lambda}{\pi} \right)^2 \sin^2\left(\frac{2\pi L}{\Lambda}\right). \tag{48}$$

When comparing (48) to (46) we see that considered first order quasi phase matching in a grating gives much larger sum frequency generation than ideal phase matching for $L >> \Lambda$.

## 11 Quantum mechanical description

In a quantum mechanical description [9, 14] sum frequency generation builds on a process where two photons generate a new photon. Such a process requires conservation of photon energy

$$\hbar\omega_3 = \hbar\omega_1 + \hbar\omega_2 \tag{49}$$

and of photon impulse

$$\hbar\beta_3 = \hbar\beta_1 + \hbar\beta_2. \tag{50}$$

Here $\hbar = h/(2\pi)$ where $h = 6.626 \times 10^{-34}$ Js is Planck's constant. We see that (49) determines the sum frequency $\omega_3$ and (50) is in agreement with the ideal phase match condition (39) derived above from coupled wave theory.

## 12 Time dependent envelopes

Having considered monochromatic waves we will now address the case where wave 1 and 2 have time dependent envelopes and generate a new wave 3 with time dependent envelope and a carrier frequency that is the sum of the two input carrier frequencies [5]. Wave 3 might have a different group velocity compared to waves 1 and 2, which then will cause a walk-off effect that we want to take into account. As in Section 8 we assume waves 1 and 2 are strong and propagate without loosing power, while wave 3 is growing slowly as a function of the distance $z$. We now introduce time dependent envelopes in (29)

$$E_k(z,t) = \frac{1}{2} \left[ A_k(z,t) e^{j\theta_k} + c.c. \right] = \frac{1}{2} \left[ E_{ck}(z,t) + c.c. \right] \tag{51}$$

for $k = 1, 2, 3$, where $E_{ck}(z,t) = A_k(z,t)e^{j\theta_k}$ are the complex fields, $A_k(z,t)$ the envelopes and $\theta_k = \omega_k t - \beta_{0k} z$. With reference to (28) and using similar arguments as in Section 8 we now focus on the equation

$$\left[\frac{\partial^2}{\partial z^2} - \mu_0\varepsilon_0\frac{\partial^2}{\partial t^2} - \mu_0\varepsilon_0\chi(t) \otimes \frac{\partial^2}{\partial t^2}\right] E_3(z,t) = \mu_0 4d\frac{\partial^2}{\partial t^2}\left[E_1(z,t)E_2(z,t)\right]. \tag{52}$$

With introduction of (51) in (52) we see that the only term on the r.h.s. that can drive a wave at $\omega_3 = \omega_1 + \omega_2$ is $E_{c1}E_{c2}$. Hence

$$\left[\frac{\partial^2}{\partial z^2} - \mu_0\varepsilon_0\frac{\partial^2}{\partial t^2} - \mu_0\varepsilon_0\chi(t) \otimes \frac{\partial^2}{\partial t^2}\right] E_{c3}(z,t) = \mu_0 2d\frac{\partial^2}{\partial t^2}\left[E_{c1}(z,t)E_{c2}(z,t)\right]. \tag{53}$$

Similarly

$$\left[\frac{\partial^2}{\partial z^2} - \mu_0\varepsilon_0\frac{\partial^2}{\partial t^2} - \mu_0\varepsilon_0\chi(t) \otimes \frac{\partial^2}{\partial t^2}\right] E_{ck}(z,t) = 0 \tag{54}$$

for $k = 1, 2$. As shown in Appendix A the equations (53) and (54) lead to the simple equation

$$\left[\frac{\partial}{\partial z} + \beta_{13}\Delta\omega\right] \tilde{A}_3(z, \Delta\omega) = \gamma d e^{j(\Delta\beta + \beta_{11}\Delta\omega)z} \tilde{A}_1 \otimes \tilde{A}_2(\Delta\omega). \tag{55}$$

where $\beta_{1k} = \frac{\partial \beta}{\partial \omega}(\omega_k)$ for $k = 1, 2, 3$, $\tilde{A}_k(\omega) = \tilde{A}_k(0, \omega)$ for $k = 1, 2$ and $\gamma$ is given by (106). From (55) we find

$$\frac{\partial}{\partial z}\left(\tilde{A}_3(z, \Delta\omega)e^{j\beta_{13}\Delta\omega z}\right) = -j\gamma d e^{j(\Delta\beta + (\beta_{13}-\beta_{11})\Delta\omega)z} \tilde{A}_1(\Delta\omega) \otimes \tilde{A}_2(\Delta\omega) \tag{56}$$

with the solution

$$\tilde{A}_3(L, \Delta\omega)e^{j\beta_{13}\Delta\omega L} = \int_0^L B(z, \Delta\omega)dz \tag{57}$$

where $B(z, \Delta\omega)$ is the r.h.s. of (56) and recalling $\tilde{A}_3(0, \Delta\omega) = 0$. Hence

$$\tilde{A}_3(L, \Delta\omega) = H(\Delta\omega)\tilde{A}_1(\Delta\omega) \otimes \tilde{A}_2(\Delta\omega) \tag{58}$$

where

$$H(\Delta\omega) = -j\gamma e^{-j\beta_{13}\Delta\omega L}\int_0^L d(z)e^{j(\Delta\beta + (\beta_{13}-\beta_{11})\Delta\omega)z}dz \tag{59}$$

is a kind of transfer function. It relates the Fourier transforms of the envelopes of the clock and signal to the Fourier transform of the output envelope at the sum frequency. It can be considered as representing a filter acting upon the spectral components of the input signal and clock and it depends on the nonlinear coefficient and the dispersive properties of the nonlinear medium in addition to other parameters. Thus $\beta_{1k} = 1/v_g(\omega_k)$ where $v_g(\omega_k)$ is the group velocity at $\omega_k$ according to equation (102). The phase $(\beta_{13} - \beta_{11})\Delta\omega z$ therefore takes into account the walk-off between signal and generated waves due to difference in group velocity. We introduce the parameter

$$\delta_v = \beta_{13} - \beta_{11} = \frac{1}{v_g(\omega_3)} - \frac{1}{v_g(\omega_1)} \tag{60}$$

as a measure of the walk-off. For bulk $LiNbO_3$ material we can calculate $\beta(\omega)$ from the relation

$$\beta(\omega) = \frac{\omega}{c} n(\omega) \tag{61}$$

where $n(\omega)$ is the refractive index. According to [15] the latter can for a temperature of 24.5 °C be approximated by the Sellmeier equation

$$n^2 = a_1 + \frac{a_2}{\lambda^2 - a_3^2} + \frac{a_4}{\lambda^2 - a_5^2} - a_6 \lambda^2 \tag{62}$$

where $\lambda$ is the wavelength in µm and $a_1 = 5.35583$, $a_2 = 0.100473$, $a_3 = 0.20692$, $a_4 = 100$, $a_5 = 11.34927$ and $a_6 = 1.5334 \cdot 10^{-2}$. A numerical calculation of

$$\beta_1 = \frac{1}{c} \frac{d(\omega n)}{d\omega} = \frac{1}{c} \left( n - \lambda \frac{dn}{d\omega} \right) \tag{63}$$

shows that $\beta_1(\lambda)$ is decreasing from 13.86 ps/mm at a wavelength of $\lambda = 0.3$ µm to a minimum of 7.262 ps/mm at $\lambda = 1.92$ µm. This means that $\delta_v = \beta_{13} - \beta_{11}$ is positive when $\omega_3$ and $\omega_1$ correspond to wavelengths in this interval.

Again, let us consider a first order grating with $l = \frac{\Lambda}{2}$ for which we have $d(z) = d_{eff} \frac{2}{\pi} \left( e^{j2\pi \frac{z}{\Lambda}} + e^{-j2\pi \frac{z}{\Lambda}} \right)$, so

$$H(\Delta\omega) = -j\gamma d_{eff} \frac{2}{\pi} e^{-j\beta_{13}\Delta\omega L} \int_0^L \left( e^{j2\pi \frac{z}{\Lambda}} + e^{-j2\pi \frac{z}{\Lambda}} \right) e^{j(\Delta\beta + \delta_v \Delta\omega)z} dz. \tag{64}$$

We now define

$$\Delta\beta_g = \frac{2\pi}{\Lambda} + \Delta\beta + \delta_v \Delta\omega \tag{65}$$

and find

$$H(\Delta\omega) \simeq -j\gamma d_{eff} \frac{2}{\pi} L e^{-j\beta_{13}\Delta\omega L} e^{j\frac{\Delta\beta_g L}{2}} \operatorname{sinc}\left( \frac{\Delta\beta_g L}{2\pi} \right) \tag{66}$$

for $|\Delta\beta_g| << \frac{4\pi}{\Lambda}$ where the contribution from the second term in the parenthesis in (64) is very small and can be ignored. For simplicity, let us assume the quasi phase match condition $\frac{2\pi}{\Lambda} + \Delta\beta = 0$ is satisfied; then $\Delta\beta_g = \delta_v \Delta\omega$ and (66) takes the form

$$H(\Delta\omega) \simeq H(0) e^{-j\frac{\Delta\omega(\beta_{13} + \beta_{11})L}{2}} \operatorname{sinc}\left( \frac{\Delta\omega \delta_v L}{2\pi} \right) \tag{67}$$

where $H(0) = -j\gamma d_{eff} \frac{2}{\pi} L \simeq -j\omega_3 \sqrt{\frac{\mu_0}{\varepsilon_3}} d_{eff} \frac{2}{\pi} L$ since $|\triangle\omega| \ll \omega_3$. The normalized spectrum defined as the norm squared of the normalized transfer function becomes

$$\left| \frac{H(\Delta\omega)}{H(0)} \right|^2 = \operatorname{sinc}^2\left( \frac{\Delta\omega \delta_v L}{2\pi} \right). \tag{68}$$

We see that $\delta_v$ controls a walk-off effect that reduces the output at the sum frequency. In the following $H(0)$ is called $H_0$.

The inverse Fourier transform of (67) is the impulse response

$$h(t) = \int_{-\infty}^{\infty} H(\omega) e^{j\omega t} df = H_0 \frac{1}{T} \Pi \left( \frac{t - \langle t \rangle}{T} \right) \tag{69}$$

where $T = \delta_v L = (\beta_{13} - \beta_{11})L$ and $\langle t \rangle = (\beta_{13} + \beta_{11})L/2$. Thus $h(t) = H_0/T$ for $\beta_{11}L < t < \beta_{13}L$ and zero elsewhere. We have assumed $\beta_{13} > \beta_{11}$. The inverse Fourier transform of (58) therefore gives the simple expression for the envelope in the time domain

$$\begin{aligned} A_3(L,t) &= h(t) \otimes (A_1(t)A_2(t)) = \int_{-\infty}^{\infty} h(t-t')A_1(t')A_2(t')dt' \\ &= H_0 \frac{1}{T} \int_{t-b}^{t-a} A_1(t')A_2(t')dt' \end{aligned} \tag{70}$$

where $a = \beta_{11}L$, $b = \beta_{13}L$ and $T = b - a$.

## 13 Signal with sinusoidal amplitude modulation and clock with higher harmonics

Let us give the mathematical foundation for the clock extraction experiment in [5] where the input signal has a sinusoidal amplitude modulation at the modulation frequency $f_{m1} = 40$ GHz. The envelope $A_1(t)$ is given by

$$A_1(t) = A_{10}[1 + m_1 \cos(\omega_{m1}t + \phi_s)] \tag{71}$$

where $A_{10}$ is the dc value, $m_1$ the modulation coefficient, $\omega_{m1} = 2\pi f_{m1}$ the angular modulation frequency and $\phi_s$ an arbitrary almost constant phase with small fluctuations. Furthermore, let us assume the local clock has an envelope $A_2(t)$ that is periodic with the fundamental frequency $f_{m2} = 10$ GHz and is given by [5]

$$A_2(t) = A_{20}\left[1 + m_2 \sum_{n=1}^{\infty} a_n \cos[n(\omega_{m2}t + \theta)]\right] \tag{72}$$

where $A_{20}$ is the dc value, $\omega_{m2} = 2\pi f_{m2}$, $m_2$ the "strength" of the modulation and $\theta$ the phase shift of the clock.

The experimental set-up is sketched in Fig. 3. The purpose was to extract a clock at 10 GHz, i.e. at the 4th sub-harmonic of the signal frequency. The 40 GHz signal was generated by a CW tunable laser followed by a Mach-Zehnder modulator driven by a 20 GHz frequency synthesizer such that the modulation frequency was doubled to 40 GHz; the signal wavelength was $\lambda_s = 1576.68$ nm. The optical clock was generated by an integrated laser and modulator driven by the VCO at the fundamental frequency $f_c = 10$ GHz ; the optical output had a 4th harmonic at $4f_c = 40$ GHz and the wavelength was $\lambda_c = 1552.97$ nm. The wavelength at the optical sum frequency was $\lambda_3 = (1/\lambda_s + 1/\lambda_c)^{-1} = 782.37$ nm suitable for the Si APD.

First we determine $A_1(t) \cdot A_2(t)$ from (71) and (72)

$$A_1(t) \cdot A_2(t) = A_{10}A_{20}\Big\{1 + m_1\cos(\omega_{m1}t + \phi_s)$$
$$+ m_2\sum_{n=1}^{\infty} a_n\cos(n\omega_{m2}t + n\theta) + \frac{m_1m_2}{2}\sum_{n=-\infty}^{\infty} a_n\cos\left[(\omega_{m1} + n\omega_{m2})t + \phi_s + n\theta\right]\Big\}$$
(73)

where $a_n = a_{-n}$ and $a_0 = 0$. Since

$$\frac{1}{T}\int_{t-b}^{t-a} \cos(\omega_{m1}t' + \phi_s)dt' = \frac{1}{T\omega_{m1}}\left[\sin(\omega_{m1}(t-a) + \phi_s) - \sin(\omega_{m1}(t-b) + \phi_s)\right]$$
$$= \mathrm{sinc}(f_{m1}T)\cos[\omega_{m1}(t - \langle t \rangle) + \phi_s]$$
(74)

the insertion of (73) in (70) leads to $A_3(L,t)$ written as a linear combination of cosine functions similar to (74) except for a dc term $H_0 A_{10} A_{20}$.

$$A_3(L,t) = A_{10}A_{20}H_0\Big\{1 + m_1\mathrm{sinc}(f_{m1}T)\cos[\omega_{m1}(t - \langle t \rangle) + \phi_s]$$
$$+ m_2\sum_{n=1}^{\infty} a_n\mathrm{sinc}(nf_{m2}T)\cos[n\omega_{m2}(t - \langle t \rangle) + n\theta]$$
$$+ \frac{m_1m_2}{2}\sum_{n=-\infty}^{\infty} a_n\mathrm{sinc}((f_{m1} + nf_{m2})T)\cos\left[(\omega_{m1} + n\omega_{m2})(t - \langle t \rangle) + \phi_s + n\theta\right]\Big\}.$$
(75)

The intensity $I(t)$ is taken as

$$I(t) = \frac{1}{2\eta_3}|A_3(L,t)|^2.$$
(76)

For the photocurrent $I_p(t)$ we only keep dc terms and terms with the lowest angular frequency $\Delta\omega_N = 2\pi\Delta f_N = \omega_{m1} - N\omega_{m2}$, where $N = 4$ [5], because the LPF in Figure 3 rejects terms at higher frequencies. So from (76) and (75) follows to 2nd order in products of $m_1$ and $m_2$

$$I_p(t) \simeq R_d A \frac{1}{2\eta_3}(A_{10}A_{20}|H_0|)^2\Big[1 + H_1 + m_1m_2a_N$$
$$\times \{\mathrm{sinc}(\Delta f_N T) + \mathrm{sinc}(f_{m1}T)\mathrm{sinc}(Nf_{m2}T)\}\cos\left[\Delta\omega_N(t - \langle t \rangle) + \phi_s - N\theta\right]\Big]$$
$$= I_{dc} + I_a\sin\epsilon(t).$$
(77)

Here $R_d$ is the responsivity, $A$ the cross sectional area of the photodiode and $H_1$ is an uninteresting constant that is a sum of terms that are of second or higher order in $m_1$ and $m_2$. We have introduced the substitutions

$$I_{dc} = R_d A \frac{1}{2\eta_3} (A_{10} A_{20} |H_0|)^2 (1 + H_1) \tag{78}$$

$$I_a = R_d A \frac{1}{2\eta_3} (A_{10} A_{20} |H_0|)^2 m_1 m_2 a_N \left(\text{sinc}(\Delta f_N T) + \text{sinc}(f_{m1} T) \text{sinc}(N f_{m2} T)\right) \tag{79}$$

and

$$\epsilon(t) = \Delta \omega_N (t - \langle t \rangle) + \phi_s - N\theta + \frac{\pi}{2}. \tag{80}$$

In Section 2 is described how a control voltage $y_d(t)$ derived in (4), here called an error signal, can tune the phase of the VCO in (3) to the same frequency as the signal. Ignoring $I_{dc}$ the error signal in the present case is $I_a \sin \epsilon(t)$ and similar to (3) it tunes the phase of the VCO according to

$$\theta(t) = \theta(0) + K_d \int_0^t I_a \sin[\epsilon(t')] dt' \tag{81}$$

where $K_d$ is a constant. The output clock from the VCO is $x_v(t) = A_v \cos[\omega_{m2} t + \theta]$ where $A_v$ is a constant. It modulates the LOL in Figure 3 and produces an envelope of the form (72). It follows from (80) and (81) that

$$\dot{\epsilon}(t) = \Delta \omega_N + \dot{\phi}_s - N\dot{\theta} = \Delta \omega_N + \dot{\phi}_s - N K_d I_a \sin[\epsilon(t)] \tag{82}$$

which is the same as (7) for $K = N K_d I_a$. Following Section 2 we now replace $K$ with $1/\tau$. When the OPLL is closed the clock $x_v(t) = A_v \cos[\omega_{m2} t + \theta]$, with $\theta$ inserted from (80), becomes

$$x_v(t) = A_v \cos \left[ \omega_{m2} t + \frac{1}{N} (\Delta \omega_N (t - \langle t \rangle) + \phi_s + \frac{\pi}{2} - \epsilon) \right] = A_v \cos \left[ \frac{1}{N} \omega_{m1} t + \psi \right] \tag{83}$$

where the phase is $\psi = (\phi_s - \Delta \omega_N \langle t \rangle + \pi/2 - \epsilon)/N$. For $t \gg \tau$ the approximate solution to (82) (see (11)) is

$$\epsilon(t) \simeq \tau \Delta \omega_N + \int_0^t e^{-(t-t')/\tau} \dot{\phi}_s(t') dt'$$
$$\simeq \tau \Delta \omega_N + \phi_s(t) - \langle \phi_s(t) \rangle \tag{84}$$

where

$$\langle \phi_s(t) \rangle = \frac{1}{\tau} \int_0^t \phi_s(t') e^{-(t-t')/\tau} dt' \tag{85}$$

is a lowpass filtered version of $\phi_s(t)$. For $\Delta \omega_N (\langle t \rangle + \tau) \ll \pi/2$ the phase $\psi(t)$ is approximately

$$\psi(t) \simeq \left( \langle \phi_s(t) \rangle + \frac{\pi}{2} \right) / N \tag{86}$$

and $x_v(t)$ is approximately

$$x_v(t) \simeq A_v \cos\left[\frac{1}{N}\left(\omega_{m1}t + \langle\phi_s(t)\rangle + \frac{\pi}{2}\right)\right]. \tag{87}$$

So the OPLL provides a clock with frequency equal to the input signal frequency divided by $N$, i.e. it provides clock recovery [5, 16]. The clock traces the signal with an averaged phase divided by $N$. Fluctuations or variations of $\phi_s(t)$ on timescales less than or of the order of $\tau$, such as high frequency phase jitter, are averaged and suppressed in $\langle\phi_s(t)\rangle$.

## 14 Experimental validation

For experimental validation, the 40 GHz signal was observed on an optical sampling scope; the scope was triggered by the recovered clock and for comparison by the frequency synthesizer driving the signal source. No noticeable difference between the two signal traces was observed thus demonstrating good quality of the recovered clock. In another experiment the error signal's sinusoidal dependence of the phase error $\epsilon$ in (77) for $\Delta\omega_N = 0$ was also confirmed; this was done by measuring the power of the error signal versus the delay, i.e. the phase difference, between the signal and the clock. A documentation of the sinusoidal phase error is given in [6] for 160 Gbit/s and 320 Gbit/s input data streams.

## 15 Error signal and clock recovery for OTDM signals up to 640 Gbit/s

In [5] also clock recovery of OTDM 160 Gbit/s, 320 Gbit/s and 640 Gbit/s has been demonstrated. The OTDM data signals were generated by a 10 GHz mode locked ERGO laser; the pulse stream was on-off modulated with a pseudo random bit stream, the pulses were compressed and by time interleaving multiplexed up to the higher bit rates mentioned. Thus the slow error signal is determined by the frequency and phase difference between the 64th harmonic of the 10 GHz clock and the signal, and as before after detection in an APD and subsequent low pass filtering the error signal is used to control the 10 GHz VCO. As a new feature compared to the previous experiment with sinusoidal signal the 10 GHz clock was used to drive another ERGO laser whose output pulses were compressed and then used as control pulses in a NOLM-demultiplexer from which the 64th received signal pulses were extracted. This 10 Gbit/s signal pulse steam was in turn used as input to a 10 Gbit/s bit error rate measurement; bit error rates below $10^{-9}$ were measured even after transmission through 50 km optical fiber thus demonstrating successful carrier recovery.

# 16 Numerical example

Let us consider two LiNbO$_3$ devices with length $L = 30$ mm and $L = 60$ mm, respectively [5]. For optical wavelengths of practical interest we have for example $\lambda_1 \simeq \lambda_2 = 1.55$ µm and $\lambda_3 = 1.55/2$ µm. For those wavelengths we have the walk-off value $\delta_v \simeq 0.30$ ps/mm. According to (68) the norm of the normalized transfer function is

$$\left| \frac{H(\Delta\omega)}{H(0)} \right| = |\text{sinc}(\Delta f \delta_v L)| \tag{88}$$

which is plotted in Figure 7 versus the frequency offset $\Delta f = f - f_3$ for the two lengths. For $L = 30$ mm we have the first zero for $\Delta f \simeq 111\,\text{GHz}$, for $L = 60$ mm half the spacing $\Delta f \simeq 55$ GHz. Furthermore, and with reference to (77), for $f_{m1} = 40$ GHz, $f_{m2} = 10$ GHz, $N = 4$, $\Delta f_N = f_{m1} - N \cdot f_{m2} \simeq 0$, $L = 60$ mm we find $\text{sinc}(\Delta f_N \delta_v L) \simeq 1$, $\text{sinc}(f_{m1}\delta_v L) = 0.34$, $\text{sinc}(Nf_{m2}\delta_v L) = 0.34$, and $\text{sinc}(f_{m1}\delta_v L) \cdot \text{sinc}(Nf_{m1}\delta_v L) = 0.12$. This shows that in the large parenthesis in (77) the sinc-product term is only about 12% of the first term.

# 17 Summary

We have given a tutorial presentation of opto-electronic clock recovery to be used in optical communication systems where the bit rate is so high that purely electronic solutions in the clock recovery process is not possible. A basic and general presentation of CR and PLL were given before focusing on optical CR and OPLL. The OPLL contains a nonlinear LiNbO$_3$ crystal in which a received signal wave and a local wave generate a third wave at the optical sum frequency. After detection of the third wave in a photodiode and subsequent low pass filtering an electrical error signal is generated determined by the frequency and phase difference between the signal and a higher harmonic of the clock. The error signal controls a VCO that delivers the electrical clock which is used to modulate the local oscillator laser that is input to the LiNbO$_3$ crystal and the clock also drives the sampling of the output signal in the optimal moments of the bits. The second harmonic process taking place in the LiNbO$_3$ crystal was described in detail in an appendix based on a nonlinear wave equation.

The theory was validated in an experiment. The purpose was to extract a clock at 10 GHz, i.e. at the 4th sub-harmonic of the signal frequency modulated at 40 GHz. The optical clock was generated by a local oscillator laser driven by the VCO at the fundamental frequency $f_c = 10$ GHZ ; the optical output had a 4th harmonic at 40 GHz. In other experiments [5] also clock recovery of OTDM 160 Gbit/s, 320 Gbit/s and 640 Gbit/s signals was demonstrated. The local optical clock was generated by a 10 GHz tunable mode locked laser, and in case of the 640 Gbit/s experiment [7] the 10 GHz pulses were also pulse compressed. The OPLL was the same as in the previous section, but it was the 64th harmonic of the clock that was used instead of the 4th. The theory for generation of the error signal for the 640 Gbit/s experiment would be a modification of the one in Section (13).

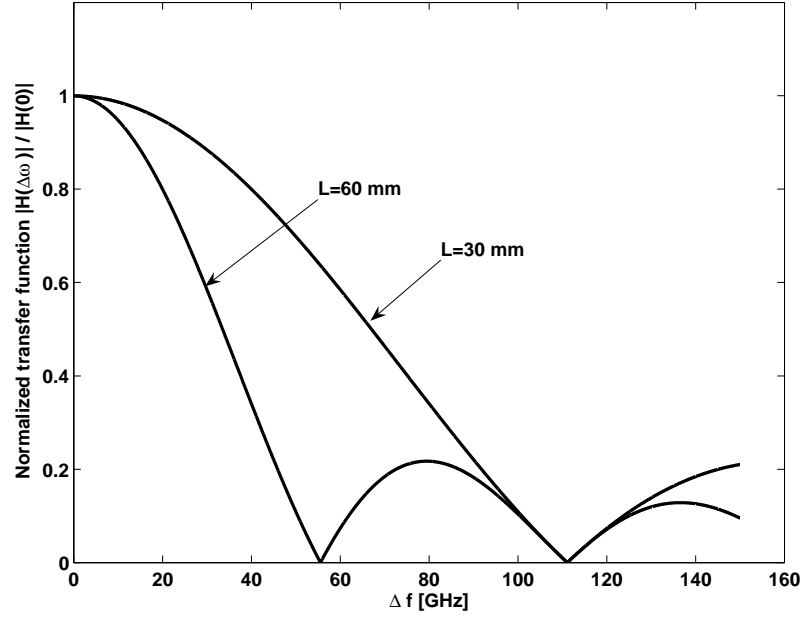

Figure 7: Normalized transfer function versus frequency offset with crystal length $L$ as parameter.

Numerical examples was given for two LiNbO$_3$ devices with length $L = 30$ mm and $L = 60$ mm, respectively [5]. For $L = 30$ mm the first zero of the normalized transfer function is for the frequency $\Delta f \simeq 111$ GHz, for $L = 60$ mm half the spacing $\Delta f \simeq 55$ GHz.

## A  Equations for the envelopes $A_k(z,t)$

For $E_{ck}(z,t) = A_k(z,t)e^{j\theta_k}$ where $\theta_k = \omega_k t - \beta_{0k} z$ the l.h.s. of (54) and (53)

$$l.h.s. = \left[\frac{\partial^2}{\partial z^2} - \frac{1}{c^2}\frac{\partial^2}{\partial t^2} - \frac{1}{c^2}\chi(t) \otimes \frac{\partial^2}{\partial t^2}\right] E_{ck}(z,t) \qquad (89)$$

can be expressed in terms of the envelopes $A_k(z,t)$ for $k = 1, 2, 3$. The first term in (89) becomes

$$\frac{\partial^2}{\partial z^2} E_{ck}(z,t) \simeq e^{j\theta_k}\left[-2j\beta_{0k}\frac{\partial}{\partial z} - \beta_{0k}^2\right] A_k(z,t) \qquad (90)$$

assuming $A_k(z,t)$ to be slowly varying with $z$ such that $\frac{\partial^2}{\partial z^2} A_k(z,t)$ can be ignored compared to $\beta_{0k}^2 A_k(z,t)$.

Introducing the Fourier transform with $\omega = 2\pi f$

$$E_{ck}(z,t) = \int_{-\infty}^{\infty} \tilde{E}_{ck}(z,\omega) e^{j\omega t} df \qquad (91)$$

we find

$$-\frac{1}{c^2} \chi(t) \otimes \frac{\partial^2}{\partial t^2} E_{ck}(z,t) = \frac{1}{c^2} \int_{-\infty}^{\infty} \chi(t') \left[ \int_{-\infty}^{\infty} \tilde{E}_{ck}(z,\omega) \omega^2 e^{j\omega(t-t')} df \right] dt'$$

$$= \frac{1}{c^2} \int_{-\infty}^{\infty} \tilde{E}_{ck}(z,\omega) \omega^2 \tilde{\chi}(\omega) e^{j\omega t} df \qquad (92)$$

and hence

$$-\frac{1}{c^2} \left[ \frac{\partial^2}{\partial t^2} + \chi(t) \otimes \frac{\partial^2}{\partial t^2} \right] E_{ck}(z,t) = \int_{-\infty}^{\infty} \frac{\omega^2}{c^2} (1 + \tilde{\chi}(\omega)) \tilde{E}_{ck}(z,\omega) e^{j\omega t} df$$

$$= \int_{-\infty}^{\infty} \tilde{E}_{ck}(z,\omega) \beta^2(\omega) e^{j\omega t} df = \int_{-\infty}^{\infty} \tilde{E}_{ck}(z,\omega) (\beta^2(\omega) - \beta_{0k}^2) e^{j\omega t} df + \beta_{0k}^2 E_{ck}(z,t) . \qquad (93)$$

By the inverse Fourier transform

$$\tilde{E}_{ck}(z,\omega) = \int_{-\infty}^{\infty} E_{ck}(z,t) e^{-j\omega t} dt = \int_{-\infty}^{\infty} A_k(z,t) e^{j(\omega_k - \omega)t} dt e^{-j\beta_{0k} z}$$

$$= \tilde{A}(z, \omega - \omega_k) e^{-j\beta_{0k} z} . \qquad (94)$$

Combining (90), (93) and (94) and using $\omega' = \omega - \omega_k$ the equation (89) then becomes

$$l.h.s. = -2j\beta_{0k} e^{j\theta_k} \frac{\partial}{\partial z} A_k(z,t) + e^{j\theta_k} \int_{-\infty}^{\infty} \tilde{A}_k(z,\omega') (\beta^2(\omega_k + \omega') - \beta_{0k}^2) e^{j\omega' t} df' . \qquad (95)$$

For slowly varying $A_k(z,t)$ the Fourier transform $\tilde{A}_k(z,\omega')$ is only nonzero for $\omega' \ll \omega_k$. In the integral in (95) we can therefore approximate $\beta^2(\omega_k + \omega') - \beta_{0k}^2$ by $2\beta_{0k}(\beta(\omega_k + \omega') - \beta_{0k}) \simeq 2\beta_{0k}\beta_{1k}\omega'$ where $\beta_{1k} = \frac{\partial \beta}{\partial \omega}(\omega_k)$. This finally gives the result

$$l.h.s. \simeq 2\beta_{0k} \left[ -j\frac{\partial}{\partial z} A_k(z,t) + \beta_{1k} \int_{-\infty}^{\infty} \tilde{A}_k(z,\omega') \omega' e^{j\omega' t} df' \right] e^{j\theta_k} \qquad (96)$$

$$= -j2\beta_{0k} e^{j\theta_k} \left[ \frac{\partial}{\partial z} + \beta_{1k} \frac{\partial}{\partial t} \right] A_k(z,t) . \qquad (97)$$

The l.h.s. in (89) is zero for $k = 1, 2$ according to (54). So for $k = 1, 2$

$$\frac{\partial}{\partial z} A_k(z, t) = -\beta_{1k} \frac{\partial}{\partial t} A_k(z, t) \tag{98}$$

with Fourier transform

$$\frac{\partial}{\partial z} \tilde{A}_k(z, \omega) = -j\beta_{1k} \omega \tilde{A}_k(z, \omega). \tag{99}$$

The solution to the latter is

$$\tilde{A}_k(z, \omega) = \tilde{A}_k(0, \omega) e^{-j\beta_{1k} z \omega} \tag{100}$$

and thus

$$A_k(z, t) = \int_{-\infty}^{\infty} \tilde{A}_k(0, \omega) e^{j\omega(t - \beta_{1k} z)} df = A_k(0, t - \beta_{1k} z) \tag{101}$$

which shows that the envelope $A_k(z, t)$ moves with the group velocity

$$v_g(\omega_k) = \frac{1}{\beta_{1k}}. \tag{102}$$

By (54) and (97) the equation for $A_3(z, t)$ becomes

$$-j2\beta_{03} e^{j\theta_3} \left[ \frac{\partial}{\partial z} + \beta_{13} \frac{\partial}{\partial t} \right] A_3(z, t) = \mu_0 2d \frac{\partial^2}{\partial t^2} \left[ E_{c1}(z, t) E_{c2}(z, t) \right] \tag{103}$$

with Fourier transform

$$-j2\beta_{03} e^{-j\beta_{03} z} \left[ \frac{\partial}{\partial z} + \beta_{13}(\omega - \omega_3) \right] \tilde{A}_3(z, \omega - \omega_3) = -\mu_0 2d\omega^2 \left[ \tilde{E}_{c1} \otimes \tilde{E}_{c2} \right] (\omega)$$

$$= -\mu_0 2d\omega^2 \int_{-\infty}^{\infty} \tilde{A}_1(z, \omega - \omega' - \omega_1) \tilde{A}_2(z, \omega' - \omega_2) df' e^{-j(\beta_{01} + \beta_{02}) z} \tag{104}$$

where we have inserted $\tilde{E}_{ck}(z, \omega) = \tilde{A}_k(z, \omega - \omega_k) e^{-j\beta_{0k} z}$ from (94). For $\Delta \omega = \omega - \omega_3$, $\Delta \beta = \beta_{03} - \beta_{01} - \beta_{02}$ and using the substitutions $\omega' = \omega_2 + \omega''$, and hence $\omega - \omega' - \omega_1 = \Delta \omega - \omega''$ for $\omega_3 = \omega_1 + \omega_2$, we find

$$\left[ \frac{\partial}{\partial z} + \beta_{13} \Delta \omega \right] \tilde{A}_3(z, \Delta \omega) = \gamma d \int_{-\infty}^{\infty} \tilde{A}_1(z, \Delta \omega - \omega'') \tilde{A}_2(z, \omega'') df'' e^{j\Delta \beta} \tag{105}$$

where

$$\gamma = \frac{\mu_0 (\omega_3 + \Delta \omega)^2}{\beta_{03}} = \eta_3 \frac{(\omega_3 + \Delta \omega)^2}{\omega_3} \tag{106}$$

in terms of the wave impedance $\eta_3$. The dependence of $\gamma$ on $\Delta\omega$ is weak and will be ignored in the following. Inserting $\tilde{A}_k(z,\omega) = \tilde{A}_k(\omega)e^{j\beta_{1k}z\omega}$ from (100) where $\tilde{A}_k(\omega) = \tilde{A}_k(0,\omega)$, and assuming $\beta_{11} \simeq \beta_{12}$, the resulting equation becomes

$$\left[\frac{\partial}{\partial z} + \beta_{13}\Delta\omega\right]\tilde{A}_3(z,\Delta\omega)$$

$$= \gamma d \int_{-\infty}^{\infty} \tilde{A}_1(\Delta\omega - \omega'')\tilde{A}_2(\omega'')e^{j[\beta_{11}(\Delta\omega-\omega'')+\beta_{12}\omega'']z}df''e^{j\Delta\beta z}$$

$$= \gamma d e^{j(\Delta\beta + \beta_{11}\Delta\omega)z} \int_{-\infty}^{\infty} \tilde{A}_1(\Delta\omega - \omega'')\tilde{A}_2(\omega'')df'' \quad (107)$$

or simply

$$\left[\frac{\partial}{\partial z} + \beta_{13}\Delta\omega\right]\tilde{A}_3(z,\Delta\omega) = \gamma d e^{j(\Delta\beta + \beta_{11}\Delta\omega)z}\tilde{A}_1 \otimes \tilde{A}_2(\Delta\omega). \quad (108)$$